\documentclass[]{aa} 
\usepackage{graphicx}
\usepackage{txfonts}

\usepackage{natbib}

\def\8p{8P/Tuttle}
\def\pdb{Plateau de Bure}

\def\meth{CH$_3$OH}

\def\oo{ON--OFF}
\def\tdeg{$^{\circ}$}
\def\mdeg{^{\circ}}

\begin{document}

   \title{Earth-based detection of the millimetric thermal emission of the
   nucleus of comet 8P/Tuttle\thanks{Based on observations carried out with the
   IRAM Plateau de Bure Interferometer. IRAM is supported by INSU/CNRS (France),
   MPG (Germany) and IGN (Spain).}
}


   \author{J. Boissier
          \inst{1,2,3}
          \and
          O. Groussin \inst{4}
          \and
          L. Jorda \inst{4}
          \and
          P. Lamy \inst{4}
          \and
          D. Bockel\'ee-Morvan \inst{5}
          \and
          J. Crovisier \inst{5}
          \and
          N. Biver \inst{5}
          \and
          P. Colom \inst{5}
          \and
          E. Lellouch \inst{5}
          \and
          R. Moreno \inst{5}
}

   \offprints{J. Boissier}

   \institute{Istituto di Radioastronomia - INAF, Via Gobetti 101, Bologna,
   Italy (e-mail: boissier@ira.inaf.it)
   \and
   ESO, Karl Schwarzschild Str. 2, 85748 Garching bei Muenchen, Germany
   \and
   Institut de radioastronomie millim\'etrique, 300 rue de la piscine, Domaine
   universitaire, 38406 Saint Martin d'H\`eres, France.
   \and
   Laboratoire d'Astrophysique de Marseille, Universit\'e de Provence, CNRS, 38
   rue Fr\'ed\'eric Joliot-Curie, 13388 Marseille Cedex 13, France
   \and
   LESIA, Observatoire de Paris, 5 place Jules Janssen, 92195 Meudon, France.
}

          
   \date{Received --; accepted --}

 
  \abstract
{Little is known about the physical properties of cometary nuclei. Apart from
space mission targets, measuring the thermal emission of a nucleus is one of the
few means to derive its size, independently of its albedo, and to constrain some
of its thermal properties. This emission is difficult to detect from Earth but
space telescopes (Infrared Space Observatory, Spitzer Space Telescope, Herschel
Space Observatory) allow reliable measurements in the infrared and the
sub-millimetre domains.} {We aim at better characterizing the thermal properties
of the nucleus of comet 8P/Tuttle using multi-wavelentgh space- and ground-based
observations, in the visible, infrared, and millimetre range. } {We used the
Plateau de Bure Interferometer to measure the millimetre thermal emission of
comet \8p at 240 GHz (1.25 mm) and analysed the observations with the shape
model derived from Hubble Space Telescope observations and the nucleus size
derived from Spitzer Space Telescope observations.} {We report on the first
detection of the millimetre thermal emission of a cometary nucleus since comet
C/1995~O1~Hale-Bopp in 1997. Using the two contact spheres shape model derived
from Hubble Space Telescope observations, we constrained the thermal properties
of the nucleus. Our millimetre observations are best match with: i) a thermal
inertia lower than $\sim$10~J~K$^{-1}$~m$^{-2}$~s$^{-1/2}$, ii) an emissivity
lower than 0.8, indicating a non-negligible contribution of the colder
sub-surface layers to the outcoming millimetre flux. }
{}

\keywords{Comet: individual: 8P/Tuttle -- Radio continuum: planetary systems --
Techniques: interferometric}

\titlerunning{Observations of 8P/Tuttle with the \pdb{} interferometer}

   \maketitle
%

\section{Introduction}

Much of the scientific interest in comets stems from their potential role in
elucidating the processes responsible for the formation and evolution of the
Solar System. They appeared in the outer regions of the protoplanetary disk,
when the giant planets were formed by core-accretion \citep{safzvj69,pol+96} or
disk instability \citep{cam78,bos03}. Together with the asteroids, Centaurs, and
transneptunian objects, comets are the remnants of the planetesimals not
accumulated into the planets. Comets formed in the giant planets region were
ejected to the outer Solar System and form the Oort Cloud. Some of them return
to the inner Solar System as long-period comets \citep[now called nearly
isotropic comets NIC, ][]{dun+88}, such as C/1995~O1~Hale-Bopp, or become
periodic comets after several perihelion passages and shortening of their orbit
due to gravitational perturbations. The later ones constitute the Halley-family
comets (or returning NIC) to which 1P/Halley and 109P/Swift-Tuttle are
significant representatives. Comet 8P/Tuttle also belongs to this group: its
orbit has a high ecliptical inclination (55\tdeg) with a short period (13.5
years). On the other hand, it is believed that the Kuiper Belt, gravitationally
influenced by the outer planets, supplied the population of Jupiter-family
short-period comets \citep[now called ecliptic comets EC, ][]{dun+88}. Comets
19P/Borrelly and 9P/Tempel~1, investigated by the Deep~Space~1 and Deep Impact
missions respectively, are typical comets of this category.

Given their different origins and the different evolution schemes followed since
their formation, one could expect to observe different physical properties for
the NICs and ECs. For the time being, no obvious correlation between the
chemical and physical properties of a comet and its dynamical class has been
measured \citep{cro+09}.

One of the possible ways to investigate differences between the two classes of
comets is to measure their size distribution. Reliable size determinations have
been obtained for only 13 NICs \citep{lam+04}, using ground- and space-based
telescopes in the visible and infrared domain. The range of radii is
surprisingly broad, from 0.4 to 37 km, much broader than that of the ECs but
with only 13 objects, robust conclusions cannot be drawn. Further measurements
are required to confirm this trend. In this context, it is important to develop
new methods to obtain reliable size estimates from the Earth. The millimetre
wavelength range is well suited for this purpose because the atmospheric
transmission is better than in the infrared, and also because the ALMA
observatory will offer unique observing capabilities in the near future.
However, the flux of a comet nucleus is much lower in the millimetre domain than
in the infrared, and observations are still challenging and reserved to a few
bright comets. When available, data sets at different wavelengths (visible,
infrared, millimetre) complete one another and allow detailed studies of the
nucleus properties.

Comet 8P/Tuttle was observed in 1992 using the Nordic Optical Telescope (NOT) at
6.3 AU from the Sun and appeared inactive at this time. From its apparent
magnitude, the nucleus diameter was estimated to 15.6~km \citep{lic+00}, but
photographic observations performed in 1980 at 2.3 AU suggested a nucleus 3
times smaller or highly elongated. The close approach to the Earth in December
2007-January 2008 (0.25 AU) and its supposed large size made it a very
interesting target for Earth based observations. In 2007-2008, the comet was
observed with different techniques in the visible \citep[lightcurve measurements
with the Hubble Space Telescope (HST),][]{lam+10}, the infrared \citep[Spitzer
Space Telescope (SST) observations,][]{grou+08,grou+10} and the radio domains
\citep[radar experiments with Arecibo,][]{har+10}. To complement this
multi-wavelength set of observations, we observed comet 8P/Tuttle with the
Plateau de Bure Interferometer to measure the millimetre thermal emission of its
nucleus. The only similar measurement of a comet nucleus was performed more than
10 years ago, in 1997 for comet Hale-Bopp \citep{alt+99}, which demonstrates the
difficutly of such observations and the unique opportunity offered by comet
8P/Tuttle.

In this paper we present the results of our observations of comet 8P/Tuttle
carried out with the Plateau de Bure interferometer at 1.25~mm (240 GHz) in
December 2007-January 2008. A description of the observations is given in
Sect.~\ref{sec-obs}. We analysed the data using a shape model of \cite{lam+10}
and a thermal model of the nucleus, which are described in Sect.~\ref{sec-mod}.
The results are presented in Sect.~\ref{sec-res} and summarized in
Sect.~\ref{sec-sum}.

\section{Observations}
\label{sec-obs}

\begin{table*}
\begin{center}
\caption{Log of the Plateau de Bure observations of the comet 8P/Tuttle}
\label{tab-obs}
\begin{tabular}{l c c c c c c}
\hline
\noalign{\smallskip}
Date& UT  & Ephemeris & $\Delta$ & r$_h$ & $\phi^a$ & Flux$^b$ \\
 & h & & AU & AU & & mJy \\
\hline
\noalign{\smallskip}
\hline
\noalign{\smallskip}
29 Dec. 2007 & 15--19 & JPL K074/19 &  0.26 & 1.11 & 54$^{\circ}$ & 2.4$\pm$0.7 \\
08 Jan. 2008 & 15--17 & JPL K074/21 &  0.28 & 1.06 & 65$^{\circ}$ & 3.0$\pm$0.5 \\
\hline
\noalign{\smallskip}
\end{tabular}
\end{center}
$^a$ Phase angle of the observations

$^b$ The flux and its uncertainty are measured by a fit of a point source to the
observed visibilities. The absolute flux calibration adds an uncertainty of
20\%.
\end{table*}

Comet 8P/Tuttle has been observed twice at a wavelength of 1.25~mm (240 GHz)
with the IRAM interferometer on 29 December 2007 and 8 January 2008
(Table~\ref{tab-obs}). The IRAM interferometer is a 6 antennas (15~m each) array
located at the \pdb{}, in the French Alps, and equipped with heterodyne, dual
polarization, receivers operating around 1, 2 and 3~mm (230, 150, and 100~GHz,
respectively). On both observing dates the array was set in a compact
configuration with baseline lengths ranging from $\sim$20 to 150~m. This results
in a synthesized beam diameter of about 1.2\arcsec{} at 240 GHz. The whole
calibration process was performed using the GILDAS software packages developed
by IRAM \citep{pet05}.

\paragraph{29 Dec. 2007:} 
For this first attempt, the comet was tracked using an ephemeris computed using
the JPL solution K074/19 for its orbital elements. The instrument was tuned in a
way to search for \meth{} lines around 241~GHz: 4 narrow correlator units were
dedicated to the \meth{} lines, and a total bandwidth of 1.2~GHz was used to
measure the continuum emission of the comet. 8P/Tuttle was observed from 15~h to
23~h UT, under correct conditions (system temperature of 200~K and a phase rms
of $\sim$30\tdeg) during the first half, until 19~h UT. The second half of the
observations suffered bad conditions (unstable phase, cloudy weather) and is not
usable. The final dataset represents $\sim$1.6~h on source aquired between 15~h
UT and 19~h UT. The gain calibration sources (observed every 22 minutes to
monitor the instrumental and atmospheric phase and amplitude variations) were
0133+476 and 0059+581. MWC349 was used to determine the absolute flux scale. The
pointing and focus corrections were measured every 40 and 80 minutes,
respectively. Some single dish spectra of the \meth{} lines were recorded all
along the observing period. The sky contribution was cancelled in position
switching mode (ON-OFF), with a OFF position distant of 5' from the comet. We do
not retrieve any \meth{} line detection in interferometric mode but the lines
were detected in the \oo{} spectra. Their analysis is presented in
\cite{biv+08acm8p}, together with other single dish observations of 8P/Tuttle.

\begin{figure}
   \centering
  \resizebox{\hsize}{!}{\includegraphics[]{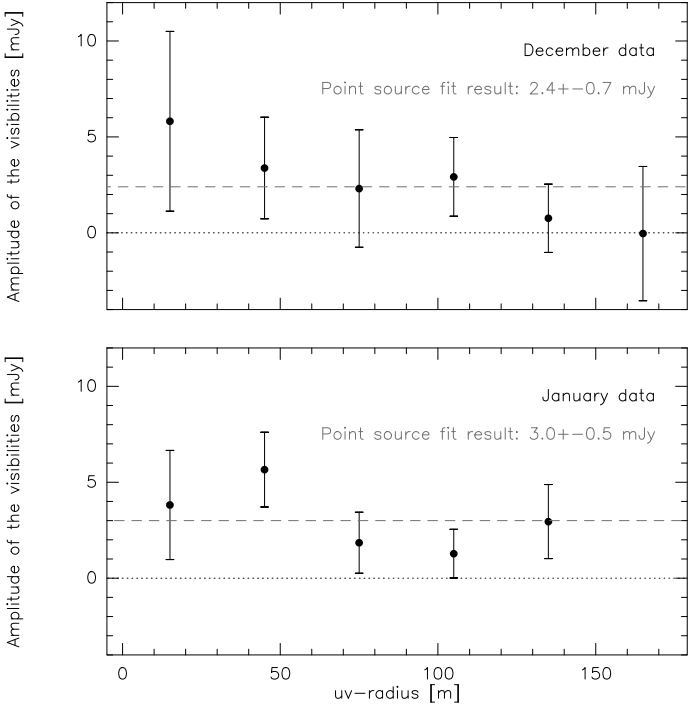}}
     \caption{Real part of the visibilities as a function of uv-radius for the
     December (top) and January data (bottom). The visibility tables have been
     shifted to the position found by the fitting procedure so that their real
     part represents the point source flux. To increase the signal to noise
     ratio on individual points, the visibilities have been averaged in 30~m
     bins of uv-radius. The result of the fitting procedure in the uv-plane is
     presented by the dashed, grey line.}
\label{fig-real}
   \end{figure}

Given the distance of the comet, the apparent diameter of a nucleus with a
radius of 10 km is about 0.1\arcsec{} and we do not expect to resolve the
nucleus of 8P/Tuttle in the Plateau de Bure data. As a result, we analysed the
continuum observations in the Fourier plane, fitting the Fourier Transform (FT)
of a point source to the observed visibilities. We measure a flux
$F=2.4\pm0.7$~mJy located at an offset of (--0.5\arcsec,--1.4\arcsec) in (RA,
Dec) with respect to the pointed position (the astrometric precision, estimated
dividing the beam size by the signal to noise ratio, is $\sim 0.3$\arcsec).
Compared to the latest ephemeris solution (JPL K074/27, including observations
performed up to october 2008) the offset is (0.4\arcsec,3.8\arcsec). To
illustrate the result of this fit, we present in Fig.~\ref{fig-real} the real
part of the visibilities as a function of the uv-radius. The visibility table
has been shifted to the position where the point source was found in the fit, so
that the imaginary part of the visibilities is null and their real part is the
source flux. On the figure, the real part of the visibilities has been averaged
in 30~m bins in uv-radius. We overplot the point source flux that was found by
the fitting procedure. The absolute flux calibration adds an uncertainty of 20\%
on both the flux and its uncertainty. We estimate the overall $\pm 1 \sigma$
limits as $F^{+1\sigma} = 1.2\times(F+1\sigma) = 3.7$~mJy and $F^{-1\sigma} =
0.8\times(F-1\sigma) = 1.4$~mJy. From this we deduce the final flux and
corresponding errorbars $F = 2.4_{-1.0}^{+1.3}$~mJy.

\paragraph{08 Jan. 2008:} 
An updated solution of orbital elements (JPL K074/21) was used to compute the
ephemeris of the comet at this date. Our single dish observations of the comet
at the IRAM-30~m telescope \citep[in early January, ][]{biv+08acm8p} showed that
its gas production was not sufficient to enable interferometric study of the
coma. As a result we dedicated all the correlator units of the \pdb{} to
continuum observations for this second run, without changing the observing
frequency 240~GHz. This results in a bandwidth of $\sim$2~GHz, which represents
a gain of a factor 1.3 in point source sensitivity with respect to the December
observations. The comet was observed from 15~h UT to 21~h UT, under poor,
degrading with time, phase stability conditions (a system temperature of 200~K
and a phase rms of $\sim$30--100\tdeg, depending on the baseline length). As a
result we use in the analysis only the 2 first hours of observations (1.3~h on
source), when the conditions were the best. Although the integration time is
longer for the December observing run, the larger bandwidth in January results
in a lower thermal noise for this dataset. The gain calibration sources were
0235+164 and 0048-097 and 3C454.3 was used as a reference to determine the
absolute flux scale.

Fitting the FT of a point source to the observed visibilities in the Fourier
plane, we retrieve a flux of 3.0$\pm$0.5~mJy located at an offset of
(1.4\arcsec,--0.1\arcsec) in (RA, Dec) with respect to the pointed position (the
astrometric precision is $\sim0.2$\arcsec). Compared to the latest solution (JPL
K074/27) the offset is (1.8\arcsec,1.2\arcsec). This result is illustrated with
the real part of the visibilities presented as a fuction of the uv-radius in
Fig.~\ref{fig-real}. Taking into account the flux calibration uncertainty, we
obtain a flux of $3.0_{-1.0}^{+1.2}$~mJy (1$\sigma$) or $3.0_{-1.8}^{+2.4}$~mJy
(3$\sigma$). In the following analysis, we only use the flux measured on January
8, as it offers a better signal to noise ratio.

We assume that the detected emission is solely due to the nucleus thermal
emission. The contribution of dust thermal emission is expected to be weak,
based on previous observations of the dust continuum in the millimetre and
submillimetre domains \citep[e.g.][]{jewluu92,jewmat97}. For exemple, using the
JCMT (HPBW = 19 \arcsec) \cite{jewmat97} measured a flux of 6.4 mJy at 350
$\mu$m for comet C/1996 B2 (Hyakutake) at $r_h$ = 1.08 AU and $\Delta =
0.12$~AU. Assuming that the dust opacity varies according to $\lambda^{-0.89}$
\citep{jewmat97} and that the dust brightness distribution varies according to
$\rho^{-1}$, where $\rho$ is the distance to nucleus, we derive a dust flux at
240 GHz of 0.05 mJy for a beam size of 1\arcsec and a geocentric distance of
0.26 AU. The gaseous activity of 8P/Tuttle in early January 2008 was typically
5--10 times lower that the activity of Hyakutake at $r_h$ = 1.08 AU
\citep{mum+96,bon+08,kob+10}. Hence, the contribution of the dust emission to
the detected flux at 240 GHz in comet 8P/Tuttle should not exceed 0.01 mJy,
compared to the 3.0~mJy of the nucleus.

\section{Nucleus thermal Model}

\label{sec-mod}

\subsection{Shape model}
\label{sec-shape}

\cite{har+08,har+10} interpreted their Arecibo radar observations of the nucleus
of comet 8P/Tuttle as implying a contact binary and proposed a shape model
composed of two spheroids in contact. \cite{lam+10} found that the light curve
of the nucleus derived from their Hubble Space Telescope observations indeed was
best explained by a binary configuration and derived a model composed of two
spheres in contact. However these two models profoundly differ first in the
direction of the rotational axis and second, in the size ratio of the two
components of the nucleus. \cite{grou+10} used both shape models to interpret
Spitzer Space Telescope thermal light curves of the nucelus and concluded that
the shape model of \cite{lam+10} better matches the SST observations.

The shape model of \cite{lam+10} consists of two contact spheres with respective
radii of 2.6$\pm$0.1~km and 1.1$\pm$0.1~km; a sphere with a radius of 2.8~km
would have the same cross-section. The pole orientation is RA~=~285\tdeg{} and
Dec~=~+20\tdeg{}, which gives an aspect angle (defined as the angle between the
spin vector and the comet-Earth vector) of 82\tdeg{} on 29 Dec. 2007 and
103\tdeg{} on 08 Jan. 2008, close to an equatorial view. The rotational period
is 11.4~h \citep{lam+08} in this model. \cite{har+10} derived a more precise
value of the period based on the radar experiments carried out at Arecibo:
11.385$\pm$0.004~h. By rotating the nucleus back from the radar epoch to the HST
epoch, we were able to connect the rotation phases and thus determine the true
or sidereal rotation period $P_{sid}$ = 11.444$\pm$0.001\ts h
\citep[see][]{lam+10}.

In view of the above discussion we used two different shape models for our
analysis: i) a simple spherical shape model with a radius of 2.8~km and ii) the
more complex and realistic shape model of \cite{lam+10}, made of two contact
spheres. In the later model, the shape is divided into 2560 triangular facets of
comparable size. We consider the size for these two shape models as robust, so
that it is a fixed parameter.

\subsection{Thermal model}
\label{sec-thermal}

The interpretation of the millimetric observations requires a thermal model for
the nucleus. We used the thermal model presented in Groussin et al. (2010) for
comet 8P/Tuttle and already extensively described in several past articles
\citep[e.g.,][]{grou+04,lam+10lut}. For each facet of the shape model, we solve
for the surface energy balance between the flux received from the Sun, the
re-radiated flux, and the heat conduction into the nucleus. As the nucleus
rotates around its spin axis, the illumination changes, and the heat conduction
equation is computed for each facet considering a one-dimensional time-dependent
equation. The projected shadows are taken into account in our model. As a
result, we obtained the temperature of each facet as a function of time, over
one rotation period. We then integrate the flux over each facet to calculate the
total millimetric flux received by the observer as a function of time and
wavelength (1.25~mm).

The active area of 8P/Tuttle is restricted to $<$15~\% of the surface
\citep{grou+10}. As for 9P/Tempel 1 with an active area of 9\% \citep{liss+05},
the sublimation of water ice can be neglected in the energy balance for the
calculation of the thermal flux emitted from the nucleus surface
\citep{grou+07di}.

\section{Results}
\label{sec-res}

Using the above simple spherical shape model with a radius of 2.8~km and our
thermal model with the parameters of \cite{grou+08} (thermal inertia
$I=0$~J~K$^{-1}$~m$^{-2}$~s$^{-1/2}$, beaming factor $\eta=0.7$, emissivity
$\epsilon=0.95$), which is in that case identical to the Standard Thermal Model
\citep{lebspe89}, we obtain a millimetric flux of 5.6~mJy at 1.25~mm on 8 Jan.
2008. This is larger than, though barely in agreement with, our observed flux of
$3.0_{-1.0}^{+1.2}$~mJy (1$\sigma$) or $3.0_{-1.8}^{+2.4}$~mJy (3$\sigma$),
which in fact corresponds to a smaller radius of $r = 2.0 \pm 0.4$~km
(1$\sigma$) using the same thermal model and parameters. In order to investigate
the origin of this discrepancy we studied the influence of several parameters in
the calculation of the flux at 1.25~mm: i) shape, ii) thermal inertia $I$, iii)
beaming factor $\eta$, and iv) emissivity $\epsilon$ from the surface and
sub-surface layers. Our goal is to investigate which of these parameters can
help to decrease the nucleus flux to make it compatible with our observations.

\subsection{Shape effect}

 \begin{figure}
   \centering
  \resizebox{\hsize}{!}{\includegraphics[]{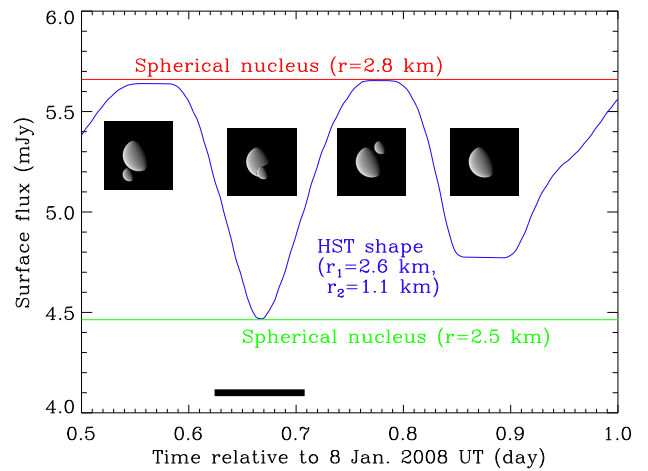}}
      \caption{Thermal lightcurve of 8P/Tuttle as seen from the Plateau de Bure
      on 8 January 2008. We used our standard set of parameters for the thermal
      inertia ($I = 0$), the beaming factor ($\eta = 0.7$) and the emissivity
      ($\epsilon = 0.95$). The shape model is that of 8P/Tuttle derived from
      Hubble Space Telescope observations \citep{lam+08,lam+10}. The lightcurve
      has been phased relatively to the radar observations of \cite{har+10},
      performed $\sim$10 rotation periods earlier. The maximum (resp. minimum)
      flux corresponds to the emission of a spherical nucleus of 2.8~km (resp.
      2.5~km). The horizontal thick solid line corresponds to the observation
      time at Plateau de Bure (2~h long).
}
         \label{fig-shape}
   \end{figure}

The simple calculation described above has been made using our spherical shape
model with a radius of 2.8~km. As described in Sect.~\ref{sec-mod}, a more
realistic shape model exists, composed of two contact spheres
\citep{lam+08,lam+10}. With this shape model, the flux is no more constant with
time as for the spherical case, but changes during the nucleus rotation to
produce a thermal lightcurve. Under our standard assumptions ($I = 0, \eta =
0.7, \epsilon = 0.95$) we calculated the synthetic thermal light curve of this
shape model, as illustrated in Fig.~\ref{fig-shape}. The flux varies with time,
with a minimum of $\sim$4.45~mJy and a maximum of $\sim$5.65~mJy. The maximum
flux is reached when the primary and secondary are both illuminated, and this is
equivalent to a spherical nucleus of radius 2.8~km as explained in
Sect.~\ref{sec-shape}. The minimum flux is reached when the secondary eclipses
the primary, and this is equivalent to a spherical nucleus of radius 2.5~km. As
the amplitude of the ligth curve is large (1.2~mJy), depending on the time of
observation, the observed flux can be quite different.

Using the sidereal rotation period of 11.444$\pm$0.001~h, we phased the
observations performed at Plateau de Bure on 8 Jan. 2008 with the Arecibo radar
experiments of \cite{har+10} performed in early January 2008. The reference
epoch of the Arecibo radar study is Jan. 4.0046, i.e. $\sim$10 rotation periods
earlier than our observations, which translates to an uncertainty of 0.0004~day,
i.e. less than 1 minute. At the reference epoch, the rotation phase is
-5$\pm$1.5\tdeg{} with 0\tdeg{} being the time when the nucleus is seen
broadside, with the larger lobe approaching \citep{har+10} and corresponds to
the maximum that follows the lightcurve minimum in Fig.~\ref{fig-shape}, on Jan.
8.75. The 1.5\tdeg{} error bar on the reference adds another 3~min of
uncertainty on the rephasing. The total uncertainty does not exceed 5~min and is
small in comparison with the length of the \pdb{} observations (2~h).

The lightcurve presented in Fig.~\ref{fig-shape} has been rephased and one can
see that the \pdb{} observations (around Jan. 8.67) were performed around its
minimum. Under the standard parameters, we expect a flux of $\sim$4.45~mJy at
that time. This is still higher than the $3.0_{-1.0}^{+1.2}$~mJy measured with
the \pdb, thus we assessed the impact of different thermal parameters of the
nucleus to investigate this discrepancy.

\subsection{Thermal inertia effect}
\label{sec-I}

\begin{figure}
   \centering
  \resizebox{\hsize}{!}{\includegraphics[]{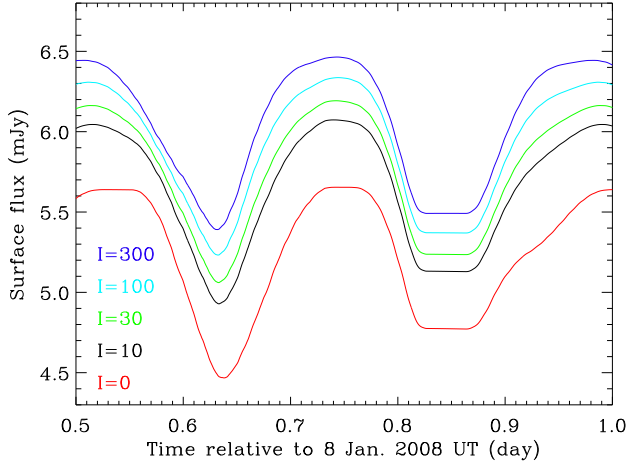}}
      \caption{ Expected thermal lightcurves of the comet 8P/Tuttle at 1.25~mm
      for different values of the nucleus thermal inertia between 0 (in red) and
      300 MKS (in blue). The beaming factor is $\eta = 0.7$ and the emissivity
      $\epsilon = 0.95$. The geometrical conditions are that of the observations
      of 8P/Tuttle on 08 January 2008 ($r_h = 1.07$~AU, $\Delta = 0.28$~AU,
      phase angle $\phi = 65\mdeg$).
}
         \label{fig-inertia}
   \end{figure}

We present in Fig.~\ref{fig-inertia} the thermal lightcurve expected from the
comet 8P/Tuttle for different values of the nucleus thermal inertia. When the
thermal inertia increases, the diurnal temperature variations decrease: the
temperatures become cooler on the day side and warmer on the night side. In our
case, since the phase angle is large (65\tdeg), the heating on the night side is
more pronounced than the cooling on the day side, and overall, the observed flux
becomes larger as thermal inertia increases. It is clear from
Fig.~\ref{fig-inertia} that a larger thermal inertia does not help to solve our
issue as the flux only gets larger. In the range
0-200~J~K$^{-1}$~m$^{-2}$~s$^{-1/2}$ for the thermal inertia estimated by
\cite{grou+08}, we then favour the lower values (typically
$\leq$10~J~K$^{-1}$~m$^{-2}$~s$^{-1/2}$).

\subsection{Beaming factor effect}

\begin{figure}
   \centering
      \resizebox{\hsize}{!}{\includegraphics[]{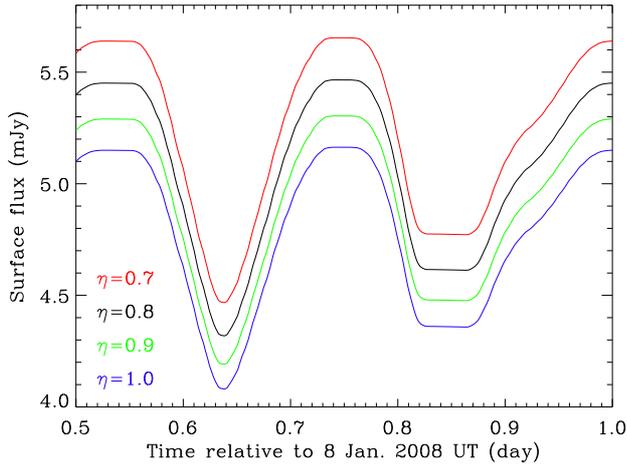}}
      \caption{Expected thermal lightcurves of the comet 8P/Tuttle at 1.25~mm
      for different values of the surface beaming factor ($\eta$). The thermal
      inertia is $I = 0$ and the emissivity $\epsilon = 0.95$. The geometrical
      conditions are identical to that of the observations of 8P/Tuttle in
      January 2008. }
         \label{fig-eta}
   \end{figure}

We present in Fig.~\ref{fig-eta} the thermal lightcurve emitted by our
synthesized, bilobate nucleus as a function of the beaming factor $\eta$. In our
thermal model, the beaming factor follows the strict definition given by
\cite{lag98} where it only reflects the influence of surface roughness and
$\eta$ must be lower or equal to 1.0. In addition, according to \cite{lag98},
$\eta$ must be larger than 0.7 to avoid unrealistic roughness, with r.m.s.
slopes exceeding 45\tdeg{}. The acceptable range for $\eta$ is then 0.7-1.0, as
illustrated in Fig.~\ref{fig-eta}. When roughness increases (i.e., $\eta$
decreases), the surface temperature increases due to self-heating on the
surface, and the total flux increases. Changing $\eta$ from 0.7 \citep{grou+08}
to 1.0 decreases the minimum flux from 4.45~mJy to 4.1~mJy, which is a small
improvement but not sufficient to explain the observed flux of 3.0~mJy.

The SST observations of comet 8P/Tuttle \citep{grou+08,grou+10} suggested a
beaming factor $\eta$ on the order of 0.7. This value is constrained by the
infrared spectrograph (IRS) observations, which lasted only 10 minutes, a short
time compared to the rotation period of $\sim$11.4~h. The SST observations
performed on 2.76 Nov. 2007 and, 141 periods later, the nucleus was in the same
rotation phase on 8.65$\pm$0.02 Jan, 2008, exactly at the time of the Plateau de
Bure observations (Fig.~\ref{fig-shape}). As a result, although higher $\eta$
values correspond to lower millimetre fluxes, we favour the $\eta$~=~0.7, which
is well constrained by the Spitzer observations.

\cite{del+03} suggested an increase of the beaming factor with phase angle for
Near-Earth Asteroids (NEAs). If confirmed, this trends could explain why our
observations performed at 65\tdeg{} phase angle favor a larger value of $\eta$,
close to one, while SST observations performed at 39\tdeg{} phase angle favor a
smaller value, close to 0.7. However, according to \cite{del+03}, their
statistic is low and more observations are required to confirm this trend.

Finally, it cannot be excluded that the effect of the roughness ($\eta$) may be
wavelength dependent. Roughness at micron scale could indeed be more important
than at millimetre scale, resulting in a smaller beaming factor in the infrared.
But this point cannot be confirmed since we cannot rule out a fractal surface,
in which case the roughness would be more or less scale-independent.

\subsection{Emissivity and depth  effect}

 \begin{figure}
   \centering
      \resizebox{\hsize}{!}{\includegraphics[]{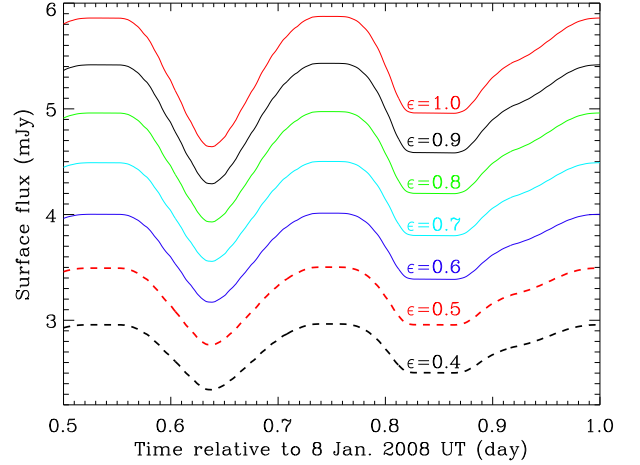}}
      \caption{Expected thermal lightcurves of the comet 8P/Tuttle at 1.25~mm
      for different values of the surface emissivity $\epsilon$ ranging from 1
      (highest, solid red line) to 0.4 (lowest, dashed yellow line). The thermal
      inertia is $I = 0$ and the beaming factor $\eta = 0.7$. The geometrical
      conditions are identical to that of the observations of 8P/Tuttle in
      January 2008. }
         \label{fig-emiss}
   \end{figure}

The thermal flux emitted from the nucleus is proportional to the surface
emissivity, as illustrated in Fig.~\ref{fig-emiss}. The flux at the lightcurve
minimum is in 1$\sigma$ agreement with our observations for an emissivity lower
than 0.8. While the surface emissivity of small bodies in the mid-infrared is
close to unity, it is still unknown at millimetre wavelengths and might be lower
than in the mid-infrared. This point has already been addressed by \cite{fer02}
for comet C/1995~O1~Hale-Bopp, where an emissivity of 0.5 is required in the
millimetic range to reconcile the nucleus sizes derived from infrared and
millimetric observations. Based on asteroid observations in the sub-millimetre,
\cite{red+98} found that the emissivity of 7 asteroids decreases with increasing
wavelength. The observations of asteroid (2867) Steins at 1.6 and 0.53~mm
performed with the radiometer MIRO onboard the Rosetta spacecraft confirm this
trend: \cite{gul+10} found an emissivity decreasing from 0.85--9 at 1.6 to
0.6--0.7 at 0.53~mm. A low millimetric emissivity of 0.6 was also found for
asteroid 4~Vesta by \cite{mullag98}. Overall, this means that our values
($\epsilon < 0.8$) are plausible.

A possible physical explanation for the lower emissivity is that a part of the
millimetre thermal flux arrises from sub-surface layers which are colder than
the surface itself. Such a temperature gradient is expected given the low
thermal intertia of the nucleus. The depth of the main contributing layer
depends on the material opacity, which is unknown. Measurements on rocks showed
that in most materials this depth lies between 3 and 100$\lambda$
\citep{camulr69}. A depth of 10$\lambda$ is commonly used when dealing with
planetary surfaces \citep[e.g. 11$\lambda$ is used for the martian surface
regolith,][]{muhber91,gol+97}. Depending on the upper layers temperature
gradient (controled by the thermal inertia) and the relative contribution of the
different layers to the overall emission, the resulting spectral emissivity of
the surface can be significantly lower than the usual infrared value close to
unity.

\begin{figure}
   \centering
      \resizebox{\hsize}{!}{\includegraphics[]{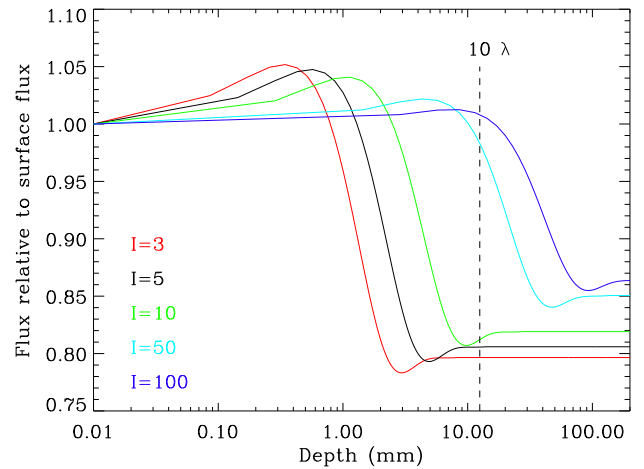}}
      \caption{Thermal emission from a sub-surface layer (relative to surface
      emission) as a function of its depth different values of the thermal
      inertia. }
         \label{fig-depth}
   \end{figure}

In order to investigate this effect, we calculated the nucleus integrated flux
as a function of depth (down to 20~cm), relative to the surface flux, for
different thermal inertia values. Figure~\ref{fig-depth} illustrates our
results: as we observed the afternoon side of the nucleus, the shadowed part
that we see is still warm because of thermal inertia and the contribution of the
first sub-surface layers is comparable or larger to the surface, down to a
certain depth (that depends on the thermal inertia) where it drops since diurnal
temperature variations become negligible.

Assuming a Poissonian weighted function (that peaks at 10$\lambda$) to describe
the relative contributions of the sub-surface layers, which seems to be a good
approximation compared to the work of \cite{mouphd}, we integrate the weigthed
flux from Fig.~\ref{fig-depth} as a function of depth, to derive the effective
outcoming flux compared to a pure surface emission. The results are presented in
Table~\ref{tab-depth}. Depending on the thermal inertia, the flux is reduced by
9\% to 17\% compared to a pure surface emission. For low thermal inertia
($\leq$10~J~K$^{-1}$~m$^{-2}$~s$^{-1/2}$), which are in better agreement with
our observations as explained in Sect.~\ref{sec-I}, the flux is reduced by
14-17\%, which corresponds to an ``effective'' emissivity of 0.79-0.82 assuming
our standard surface emissivity of 0.95. According to Fig.~\ref{fig-emiss} and
as explained above, this allows to match our observations within 1$\sigma$.

Our results confirm that a low emissivity in the millimetric wavelength range
can result from the emission of sub-surface layers with a low thermal inertia.
In the mid-infrared, typically around 10~$\mu$m, 10$\lambda$ corresponds to
0.1~mm, a depth over which the flux is close the surface flux
(Fig.~\ref{fig-depth}), explaining why the above effect only matters in the
millimetric wavelength range and beyond, in the centimetric and radar wavelength
range. However, for the centimetric and radar, there exists a few notable
exceptions, and in particular Ganymede, which centrimetric emission corresponds
to a brightness temperature of about 55-88 K, below any acceptable sub-surface
temperature for this body \citep{muhber91}, and whose radar reflection is highly
anomalous \citep{ostsho90}. In this case, a higher reflectivity must also be
invoked, likely resulting from backscattering of the surface.

\begin{table}
\begin{center}
\caption{Nucleus thermal flux integrated over depth, compared to pure surface emission (1.0).}
\label{tab-depth}
\begin{tabular}{c c c}
\hline
\noalign{\smallskip}
Thermal inertia & Relative depth-integrated flux  \\
J~K$^{-1}$~m$^{-2}$~s$^{-1/2}$ &  \\
\hline
\noalign{\smallskip}
\hline
\noalign{\smallskip}
3   & 0.83 \\
5   & 0.84 \\
10  & 0.86 \\
50  & 0.88 \\
100 & 0.91 \\
\hline
\noalign{\smallskip}
\end{tabular}
\end{center}
\end{table}

\label{sec-depth}

\section{Summary}
\label{sec-sum}

 \begin{figure}
   \centering
  \resizebox{\hsize}{!}{\includegraphics[]{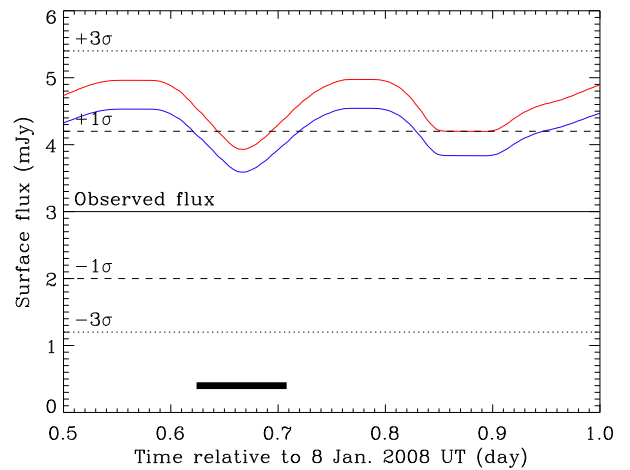}}
      \caption{ Same Figure as Fig.~\ref{fig-shape} but for $I$~=~0,
      $\epsilon$~=~0.8, $\eta$~=~0.7 (red curve) or $\eta$~=~1.0 (blue curve).
      For $\eta$~=~1.0, which is our prefered value, about half of the thermal
      lightcurve agrees with our observations at the 1$\sigma$ level.}
         \label{fig-final}
   \end{figure}

We report here the first detection of the thermal emission of a cometary nucleus
at millimetre wavelength since C/1995~O1~Hale-Bopp, performed at Plateau de Bure
on 8 Jan. 2008. Using the rotation period derived by \cite{har+10}, we phased
our observations with respect to their radar experiments and found that \pdb{}
observations were performed at a minimum of the lightcurve. We used the shape
model developped by \cite{lam+10} for 8P/Tuttle nucleus to compute thermal
emission lightcurves using different thermal parameter sets. From our analysis,
we can conclude that to match our observed flux of $3.0_{-1.0}^{+1.2}$~mJy at
1.25~mm, the thermal inertia of the nucleus should be low, typically
$\leq$10~J~K$^{-1}$~m$^{-2}$~s$^{-1/2}$, in agreement with the range
0-200~J~K$^{-1}$~m$^{-2}$~s$^{-1/2}$ derived from Spitzer observations
\citep{grou+08}. In addition, the millimetric emissivity of the nucleus should
be lower than 0.8. According to our thermal model, such a low value can be due
to the non-negligible emission of the sub-surface layers of the nucleus, which
are colder than the surface at depth of a few millimetre or more for low thermal
inertia values. Similar and even lower values of the millimetric emissivity have
been already mentionned to explain the low flux emitted by asteroids
\citep{red+98,mullag98,fer02}. As to the beaming factor, the lowest flux of the
lightucurve is in 1$\sigma$ agreement with our observed flux for any value in
the plausible range 0.7--1.0, with a better agreement towards higher edge.
However we favour $\eta$~=~0.7, the only value in agreement with Spitzer (IRS)
observations \citep{grou+10}. From the above conclusions, we generated a
synthetic lightcurve for the nucleus of comet 8P/Tuttle, as observed from the
Plateau de Bure on 8 Jan. 2008, using the following parameters: $I$~=~0,
$\epsilon$~=~0.8 and $\eta$ in the range 0.7-1.0 to cover the possible values.
The result is illustrated in Figure~\ref{fig-final}.

In the future, the Rosetta spacecraft, equipped with a microwave instrument
operating at 190 and 562 GHz \citep{gul+07}, will provide an unequaled dataset
to study the surface of a cometary nucleus and its thermal properties. It will
be also important to repeat ground-based studies on other comets, to determine
whether all comets present similar properties or if they change from one comet
to another, revealing different surface natures. The (sub-)millimetre
interferometer ALMA will offer a significant gain in sensitivity and allow
similar studies on a larger sample of comets, from both the ecliptic and the
Oort cloud dynamical classes.

\begin{acknowledgements}The research leading to these results has received
funding from the European Community's Seventh Framework Programme
(FP7/2007--2013) under grant agreement No. 229517 \end{acknowledgements}


\bibliographystyle{aa}
\bibliography{mnemonic,bigbiblio}

\end{document}